\documentclass{article}
\usepackage[T1]{fontenc} % add special characters (e.g., umlaute)
\usepackage[utf8]{inputenc} % set utf-8 as default input encoding
\usepackage{ismir,amsmath,cite,url}
\usepackage{graphicx}
\usepackage{color}
\usepackage{multirow} 
\usepackage{lineno}
\title{Lyrically Speaking: Exploring the Link Between Lyrical Emotions, Themes and Depression Risk}

\multauthor
{Pavani Chowdary$^1$ \hspace{1cm} Bhavyajeet Singh$^1$ \hspace{1cm} Rajat Agarwal$^1$} { \bfseries{Vinoo Alluri$^1$ \hspace{1cm}}\\
 $^1$ Music Cognition Group, International Institute of Information Technology, Hyderabad, India\\
{\tt\small \{boddu.pavani, bhavyajeet.singh, rajat.agarwal\}@research.iiit.ac.in, vinoo.alluri@iiit.ac.in}
}

\def\authorname{P. Chowdary, B. Singh, R. Agarwal and V. Alluri}

\begin{document}

\maketitle
\begin{abstract}
Lyrics play a crucial role in affecting and reinforcing emotional states by providing meaning and emotional connotations that interact with the acoustic properties of the music. Specific lyrical themes and emotions may intensify existing negative states in listeners and may lead to undesirable outcomes, especially in listeners with mood disorders such as depression. Hence, it is important for such individuals to be mindful of their listening strategies. In this study, we examine online music consumption of individuals at risk of depression in light of lyrical themes and emotions. Lyrics obtained from the listening histories of 541 Last.fm users, divided into At-Risk and No-Risk based on their mental well-being scores, were analyzed using natural language processing techniques. Statistical analyses of the results revealed that individuals at risk for depression prefer songs with lyrics associated with low valence and low arousal. Additionally, lyrics associated with themes of \textit{denial, self-reference} and  \textit{blame} were preferred.  This study opens up the possibility of an approach to assessing depression risk from the digital footprint of individuals and potentially developing personalized recommendation systems. 

\end{abstract}
\section{Introduction}

Depression is one of the leading causes of disability in young adults globally, according to the World Health Organization \cite{depression2017other}. It has the potential to hinder and curb development in personal and social avenues of life, making it a debilitating condition. This underscores the imperative to identify and address it in the early stages.

 % Music is often associated with emotional experiences [Bosacki, O'Neill 2015]. 
 Music plays an important role in regulating mood and emotions \cite{north2004uses}. Musical preferences and music listening habits are known to invoke and reinforce moods and emotions and satisfy psychological needs \cite{doi:10.1177/0305735616663313, schafer2016goals}. Emotionally vulnerable young adults were found to have more intense relationships with music\cite{10.1093/acprof:oso/9780198530329.003.0007}. An increased emotional reliance on music was also observed in such individuals\cite{doi:10.1177/0044118X13501343}. However, certain music engagement behaviors and strategies are associated with indicators of poor mental health and do not always lead to the alleviation of existing depressive symptoms \cite{10.3389/fpsyg.2019.01199}. Individuals who are depressed or at risk of depression are often unconscious of using music as a tool to improve emotional states \cite{mcf}, which might lead to adverse outcomes. This highlights the importance of addressing and studying music listening behaviors of individuals prone to depression risk for developing intervention methods to come up with listening strategies that may lead to positive outcomes.  

\begin{figure*}[hbt!]
 \centerline{{
 \includegraphics[alt={Three components: Data Collection, Lyrics Processing, Statistical Analysis. The Data Collection section contains Last.fm users, Mental Well-being Survey and Music Listening History. The lyrics from the music listening histories are used for Emotion Quadrant Extraction and Theme Extraction in the Lyrics Processing section. The mean Quadrant Prevalence Scores and mean theme Frequency Scores obtained there, combined with user categorization based on K-10 scores, are used for Emotion-based Analysis and Theme-based Analysis in the Statistical Analysis Section.},width=1.9\columnwidth]{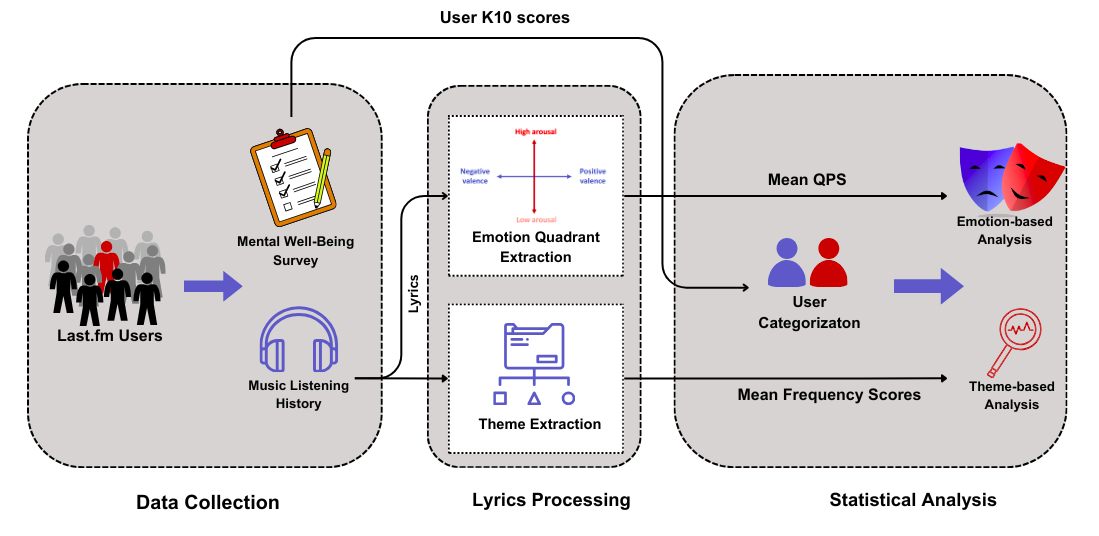}}}
 \caption{\textit{Methodology}}
 \label{fig:methodology_fig}
\end{figure*}

Online music streaming platforms such as Spotify\footnote{www.spotify.com}, Last.fm\footnote{www.last.fm}, and Apple Music\footnote{music.apple.com} offer their users a large variety of songs across genres and make it possible to study the musical digital footprints of their users.  Last.fm allows the extraction of the listening histories of its users and the corresponding metadata, which prompted several studies that utilized Last.fm to study naturally occurring user listening behaviors in light of depression risk \cite{surana2020tag2risk, Surana_2020, Shriram_2021}.  However, the relationship between lyrics, specifically the semantics and emotional connotations of lyrics, and depression has received little to no attention in the literature, while lyrics were found to play a crucial role in affecting emotional states \cite{Demetriou2018VocalsIM}.  Lyrics were also found to be essential for depicting \textit{sadness} in music \cite{Brattico2011AFM}. This study seeks to address this gap by examining the relationship between lyrical emotions and themes extracted from user listening histories and depression risk.

\section{Background and Related Work}

We highlight previous studies that used online music listening histories from Last.fm and preferences to identify different trends and characteristics in music listening behaviors of individuals at risk of depression. Surana et al.\cite{surana2020tag2risk} was the first such study, which used user-annotated tags from Last.fm, to identify emotion- and genre-tag preferences of individuals at risk of depression. The results of this study revealed that At-Risk individuals consume music that is tagged with emotions representing sadness, such as \textit{sad, depressed, dead, low} and \textit{miserable}, and belonging to genres such as \textit{neo-psychedelic, dream-pop} and \textit{indiepop}. In a later study, Surana et al. \cite{Surana_2020} studied emotions in relation to acoustic features of music as dynamic measures across the span of six months by dividing the listening history into sessions based on periods of inactivity to observe any dynamic patterns in music listening behaviors of At-Risk individuals. This study found that individuals at risk of depression rely more heavily on music and tend to listen to the same songs repeatedly. It was also found that they tend to listen to sad music for longer periods.

% based on periods of inactivity. 

Shriram et al. \cite{Shriram_2021} were the first to study lyrics in this context in terms of lyrical repetitiveness and compressibility. The results revealed that At-Risk individuals prefer music with lower lyrical simplicity (lower compressibility) and greater information content, especially for music that is characterized as sad. 

However, no study to date has explored the link between lyrical emotions and themes extracted from online listening histories and depression risk, to the best of our knowledge. 
This link is crucial to investigate because lyrics reinforce negative states and, in dire situations, lead to maladaptive outcomes.
The only study that has looked into lyrical themes associated with maladaptive listening strategies, which is known to be a proxy for depression risk \cite{hums}, was done by Singh et al. \cite{singh}. They explored the link between lyrical themes extracted using DICTION\footnote{www.dictionsoftware.com} and unhealthy music engagement strategies characterized by the Unhealthy-Healthy music scale (HUMS) \cite{hums}, which indirectly indicates depression risk. This study revealed that individuals who engage in unhealthy and maladaptive listening strategies listen to music with lyrical themes representing \textit{self-reference} and \textit{blame}. However, this has been done in the context of online discourse surrounding depression on Reddit. This raises the question of whether similar behavior can be observed in the lyrical content derived from the listening histories of individuals at risk on music streaming platforms. In this study, we investigated the relationship between individuals' online listening histories and their risk of depression in the context of lyrical emotions and themes.

 Based on previous research in the field \cite{surana2020tag2risk, Surana_2020, Shriram_2021, singh}, we hypothesize the following:

\begin{itemize}
    \item Building on prior research demonstrating a preference for \textit{sad} music among At-Risk individuals  \cite{surana2020tag2risk, Surana_2020}, we hypothesize that these individuals exhibit a greater preference for music with lyrics associated with low valence and low arousal.

    \item In line with the established link between lyrical themes and unhealthy music engagement behaviors shown by Singh et al.  \cite{singh}, we hypothesize that At-Risk individuals consume music that is higher in terms of themes such as \textit{self-reference}, and \textit{blame}. 
    % which was in turn associated with depression risk,
    
\end{itemize}

\section{Methodology}\label{sec:methodology}

Figure \ref{fig:methodology_fig} summarizes the procedure used in our study, which is described as follows.

\subsection{Dataset}

We used the dataset from Surana et al. \cite{surana2020tag2risk, Surana_2020}
for our analysis. The dataset consists of the six-month music listening history of 541 Last.fm users (Mean Age = 25.4, SD = 7.3), of which 444 were male, 82 were female, and 15 identified as other. 
This data was acquired by means of a survey that was posted on Reddit\footnote{www.reddit.com} and Facebook\footnote{www.facebook.com} Last.fm pages. Informed consent was taken from the participants and participation was completely voluntary. They were informed that the study posed no risks and that their confidentiality would be maintained. The analysis was performed at a group level, ensuring no individuals could be identified. 

\subsubsection{Measure of Depression Risk}
To assess mental well-being, Kessler's Psychological Distress Scale (K10) questionnaire \cite{kslr} is utilized, which measures psychological distress with a focus on symptoms of anxiety and depression. Following the approach of Surana et al. \cite{surana2020tag2risk}, participants scoring 29 or higher on the K10 questionnaire were classified as being in the “At-Risk” group for depression, while those scoring below 20 constitute the “No-Risk” group. Out of the total users, 193 individuals were in the No-Risk group, and 142 were in the At-Risk group.

\subsubsection{Listening History and Lyrics}
Lyrics for the tracks in the listening histories are extracted from Genius.com and MetroLyrics.com. Lyrics for approximately 76\%  of the entire repository of songs were obtained. Songs with no lyrics comprised around 4\% of the dataset. 

\subsection{Lyrics Processing}

\subsubsection{Lyrics-Emotion Mapping}

\begin{figure}[!hbt]
 \centerline{{
 \includegraphics[alt={The figure depicts the two-dimensional valence-arousal space, with four quadrants representing the emotions of happiness, anger, sadness and tenderness respectively.}, width=0.9\columnwidth]{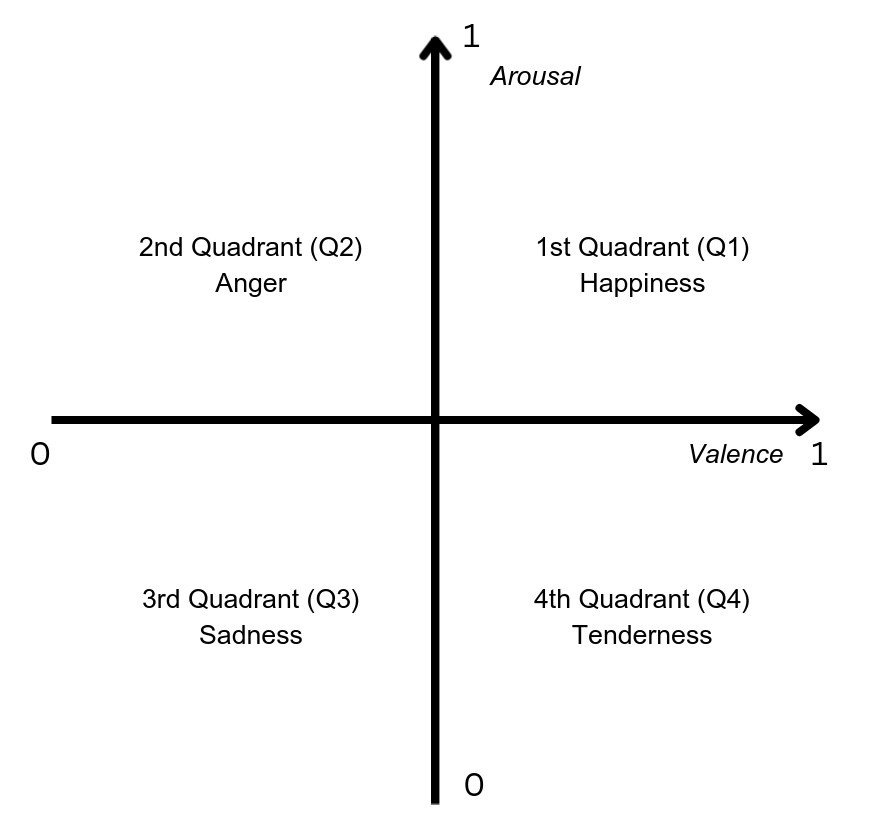}}}
 \caption{\textit{Two-dimensional Valence-Arousal space}}
 \label{fig:va_space}
\end{figure}

Lyrics were projected onto the Russell's Complex Model of Affect \cite{article}, which is used to organize emotions along two orthogonal dimensions: \textit{Valence}, which represents pleasantness and \textit{Arousal}, which represents energy. As can be seen in Figure \ref{fig:va_space}, The first quadrant represents high valence and high arousal (happiness), the second quadrant represents low valence and high arousal (anger), the third quadrant represents low valence and low arousal (sadness), and the fourth quadrant represents high valence and low arousal (tenderness). We employed a model proposed in Agarwal et al. \cite{Agrawal_2021} to map each song's lyrical content to a quadrant in the VA space. The architecture of the model is a deep neural network architecture that employs XLNet \cite{yang2020xlnet}, an advanced bidirectional transformer, to perform multitask learning for emotional classification based on song lyrics. The model is trained on the MoodyLyrics \cite{ccano2017moodylyrics} and MER \cite{Malheiro2018EmotionallyRelevantFF} datasets, which consist of songs uniformly distributed across the four quadrants of the Russell's Valence-Arousal circumplex model. The model was used to return a single quadrant label for each song by mapping the song's lyrical content to the 2D Valence-Arousal space, which was used to categorize the tracks into one of the four quadrants.

To quantify user preferences for quadrant categories, we computed a quadrant prevalence (QPS) score for each quadrant, which is determined by the proportion of tracks from each user's listening history, within the respective quadrants, as shown in Equation \ref{eq:1}. The top 100 most frequently listened songs were identified and assigned weights based on listening frequency. The QPS for each user and quadrant was then calculated as the average weighted frequency of songs belonging to that specific quadrant in their listening history.

\begin{equation}
\label{eq:1}
    QPS(u_j, q_k) = \frac{\sum_{s_i \in L(u_j)} w(s_i) \cdot I (q(s_i) = q_k)} {\sum_{s_i \in L(u_j)} w(s_i)}
\end{equation}
\\
where, \\
$QPS(u_j, q_k)$ : QPS of quadrant $q_k$ for user $u_j$ \\
$s_i$ : a song in the top 100 songs of the user's listening history $L(u_j)$ \\
$w(s_i)$ : the weight (listening frequency) of song $s_i$. \\ 
$q(s_i)$ : quadrant assigned to song $s_i$ \\
$I(.)$ : indicator function that equals 1 if the condition inside is true (song $s_i$ belongs to quadrant $q_k$) and 0 otherwise.

\subsubsection{Lyrics-Semantic Themes Mapping}
Following a similar approach to Singh et al. \cite{singh}, we used DICTION to analyze the lyrics and identify underlying semantic themes. DICTION operates through the use of dictionaries that contain lists of words associated with specific linguistic, emotional and cognitive contexts. There are 5 themes and 35 sub-themes in total. We chose the themes \textit{Self-reference, Blame, Optimism, Hardship, Satisfaction, Inspiration, Exclusion} and \textit{Denial}. The frequency scores corresponding to all the themes and sub-themes for each song were obtained, based on the occurrence of the words from the lyrics in the dictionary lists. Similar to the quadrant prevalence scores, mean frequency scores (MFS) were computed for all the themes per user, as shown in Equation \ref{eq:2}.

\begin{equation}
\label{eq:2}
    MFS(u_j, t_k) = \frac{ \sum_{s_i \in L(u_j)} w(s_i) \cdot T_k(s_i)} { \sum_{s_i \in L(u_j)} w(s_i) }
    % \label{equation:mfs}
\end{equation}
\\
where, \\ 
$MFS(u_j, t_k)$ : MFS of theme $t_k$ for user $u_j$ \\
$s_i$ : a song in the top 100 songs of listening history $L(u_j)$ \\
$w(s_i)$ : the weight (listening frequency) of song $s_i$. \\ 
$T_k(s_i)$ : the frequency score assigned to song $s_i$ for theme $t_k$.

\subsection{Statistical Testing}

We divided the users into At-Risk and No-Risk groups based on their K10 scores, as mentioned before. For each quadrant, we performed a two-tailed Mann-Whitney U (MWU) test on the quadrant prevalence score between the At-Risk and No-Risk groups. Similarly, a MWU test was performed for the mean frequency score between At-Risk and No-Risk groups, for each theme selected.

\section{Statistical Testing and Results}\label{sec:results}

\subsection{Lyrical Emotion-Based Results}

\begin{figure}[hbt!]
 \centerline{{ \includegraphics[alt={Violin plot of mean Quadrant Prevalence Scores for At-Risk and No-Risk groups. The plot serves as a comparison of medians in the cases of quadrant 1, quadrant 2, quadrant 3 and quadrant 4, in between At-Risk and No-Risk groups. The plots show that the median corresponding to the At-Risk group is greater than the one corresponding to the No-Risk group. There is no significant difference in the case of the other three quadrants.}, width=0.9\columnwidth]{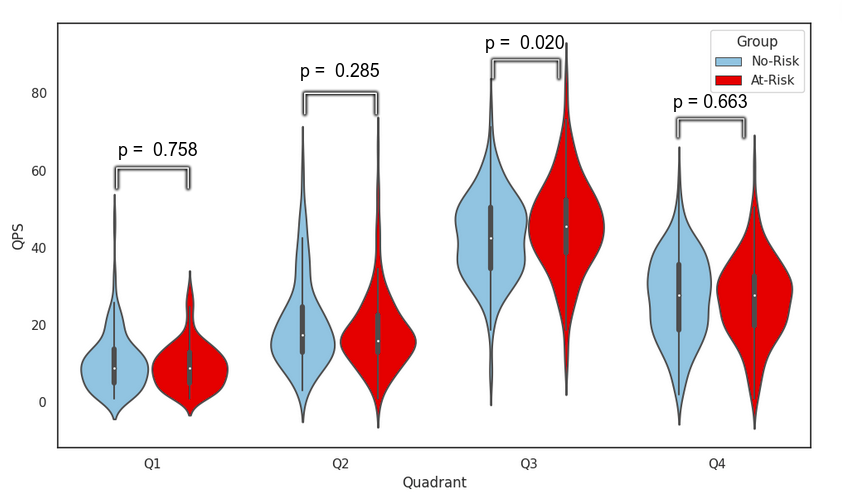}}}
 \caption{\textit{Violin plots of mean QPS per quadrant for At-Risk and No-Risk groups}}
 \label{fig:violin}
\end{figure}

A statistically significant difference (p < 0.05) was observed for the QPS corresponding to Q3, which is representative of low valence and low arousal values (U-statistic = 15727.5, p = 0.02), with a higher median for the At-Risk group, as can be seen in \ref{fig:violin}. We observed no significant differences between the distributions of the At-Risk and No-Risk groups for the other quadrants. The same trend was observed when the top 250 songs were considered.

\subsection{Lyrical Theme-based Results}

We found significant differences (p<0.05) between At-Risk and No-Risk groups in the case of \textit{denial, self-reference} and \textit{blame}. As can be inferred from Table 1, The median for the At-Risk group is higher in the cases of the themes \textit{denial, self-reference} and \textit{blame}.

\begin{table}[!hbt]
\centering
\begin{tabular}{| l | l | l | l |}
\hline
\multirow{6}{5em}{At-Risk > No-Risk} & Theme& U-statistic & p-value\\
\hline
& Denial& 15950.0& 0.010\\

& Self-reference & 15968.0 & 0.009 \\

& Blame & 15691.5 & 0.023 \\

\hline
\end{tabular}
\caption{\textit{MWU Test results for the Mean Frequency Scores between the At-Risk and No-Risk groups; Here At-Risk > No-Risk refers to the themes where the median is greater in the case of the At-Risk group}}

\end{table}
% \end{table}

\section{Discussion}

This study is the first of its kind to explore the association between risk for depression and the emotional and thematic connotations of the lyrical content of the music individuals engage with online, as opposed to studies in lab settings or self-reported data. Our results are in concordance with our initial hypotheses, in addition to revealing novel findings.

The At-Risk group exhibited a higher median score for Q3 (low valence, low arousal) than the No-Risk group. This finding suggests a greater prevalence of sadness-related lyrical content in the music listened to by the At-Risk group. The stronger association of the At-Risk group with sadness aligns with the results of past research studies on the topic. These results suggest that At-Risk individuals tend to consume music with lyrics that reflect their negative emotional states.

As hypothesized, the themes \textit{self-reference} and \textit {blame} were more prevalent in the At-Risk group. Additionally, the themes \textit{denial} was also shown to be preferred by At-Risk individuals. Since the themes of \textit{blame} and \textit{self-reference} in lyrics were found to be associated with unhealthy listening strategies \cite{singh}, listening to music associated with these themes would not be beneficial to individuals at risk of depression, and in some cases may lead to negative outcomes. This highlights the importance of mindful consumption strategies for music for people at risk of depression.

In conclusion, our results show that certain music engagement strategies are maladaptive in nature and should be avoided to prevent the worsening of mood, building on top of previous studies in the area. These results can potentially aid in developing intervention strategies based on lyrical content that should be avoided for better outcomes from music listening.

\subsection{Limitations}

A limitation of this study is the exclusive focus on lyrical emotions and themes. The interaction of these with the acoustic properties of music and lyrical complexity, as well as in the context of depression risk, could be explored, which could possibly yield a better understanding. The interaction between lyrical themes and emotional connotations is also something that is yet to be studied. Another limitation is that this study exclusively focuses on music with English lyrics. We have also used a predetermined set of themes offered by DICTION, which may not be enough to capture several lyrical themes. An approach to solve this would be to use Large Language Models (LLMs) to generate themes from a repository of songs and then using them for the scoring.

\subsection{Future Work}

The results from this paper can be used in building music recommendation systems for depressed individuals that tailor the recommendations, keeping in mind the emotions and themes that are associated with mood worsening and maladaptive behaviors to maximize the positive outcomes through music listening. These results could be combined with other measures associated with such behaviors. This work also opens up the possibility of early depression risk prediction from online music listening behaviors, in terms of lyrics, by cementing lyrical emotions and themes as indicators for depression risk. These measures, in addition to acoustic features \cite{Surana_2020} and other indicators such as lyrical complexity and social tags \cite{Shriram_2021, surana2020tag2risk}, could potentially be used to develop a multi-modal depression risk prediction system.

\bibliography{ISMIRtemplate}

\end{document}